# Single Source - All Sinks Max Flows in Planar Digraphs

Jakub Łącki[*]    Yahav Nussbaum[†]    Piotr Sankowski[‡]    Christian Wulff-Nilsen[§]

**Abstract**

Let $G = (V, E)$ be a planar $n$-vertex digraph. Consider the problem of computing max $st$-flow values in $G$ from a fixed source $s$ to all sinks $t \in V \setminus \{s\}$. We show how to solve this problem in near-linear $O(n \log^3 n)$ time. Previously, no better solution was known than running a single-source single-sink max flow algorithm $n-1$ times, giving a total time bound of $O(n^2 \log n)$ with the algorithm of Borradaile and Klein.

An important implication is that all-pairs max $st$-flow values in $G$ can be computed in near-quadratic time. This is close to optimal as the output size is $\Theta(n^2)$. We give a quadratic lower bound on the number of distinct max flow values and an $\Omega(n^3)$ lower bound for the total size of all min cut-sets. This distinguishes the problem from the undirected case where the number of distinct max flow values is $O(n)$.

Previous to our result, no algorithm which could solve the all-pairs max flow values problem faster than the time of $\Theta(n^2)$ max-flow computations for every planar digraph was known.

This result is accompanied with a data structure that reports min cut-sets. For fixed $s$ and all $t$, after $O(n^{3/2} \log^{3/2} n)$ preprocessing time, it can report the set of arcs $C$ crossing a min $st$-cut in time roughly proportional to the size of $C$.

[*]Institute of Informatics, University of Warsaw, `j.lacki@mimuw.edu.pl`.

[†]The Blavatnik School of Computer Science, Tel Aviv University, `yahav.nussbaum@cs.tau.ac.il`.

[‡]Institute of Informatics, University of Warsaw and Department of Computer and System Science, Sapienza University of Rome, `sank@mimuw.edu.pl`.

[§]Department of Computer Science, University of Copenhagen, `koolooz@diku.dk`.

# 1 Introduction

Computing max flow in a graph is a classical algorithmic problem dating back to Ford and Fulkerson [13, 14]. Given a graph $G = (V, E)$ with arc capacities, a source $s \in V$ and a sink $t \in V \setminus \{s\}$, the problem is to send as much flow as possible from $s$ to $t$ without violating capacity constraints and flow conservation at vertices in $V \setminus \{s, t\}$. Many polynomial-time algorithms for this problem are known. In this paper we focus on planar graphs. There exists an $O(n \log n)$ time algorithm for planar digraphs [3, 10] and an $O(n \log \log n)$ time algorithm for planar undirected graphs [21]. More recently, an $O(n \log^3 n)$ time algorithm for a more general problem with a set of sources and a set of sinks was proposed [4].

It is natural to consider the problem of finding multiple max flows and min cuts, e.g., to study the connectivity of the graph. For undirected (not necessarily planar) graphs, min $st$-cuts for all source/sink pairs $(s, t)$ form a laminar family and can be compactly represented by a tree structure known as a *Gomory-Hu tree* [16]. Finding a Gomory-Hu tree reduces to solving $n - 1$ min $st$-cut problems [17]. For planar undirected graphs, this gives an $O(n^2 \log \log n)$ time bound using the algorithm in [21]. This time bound was significantly improved to $O(n \log^5 n)$ in [6]. In particular, this means that all max $st$-flow/min $st$-cut values can be computed in near-linear time.

We consider the same problem but for planar *digraphs*. This problem is more challenging since the Gomory-Hu property no longer holds, even for planar digraphs [2]. In fact, there is an example of (non-planar) digraphs with $\Theta(n^2)$ distinct max flow values [22]. We show that this bound holds for planar digraphs as well. Figure 1 shows a plane $n$-vertex digraph with $\Theta(n^2)$ different cycles. There is a well-known duality between shortest cycles in the primal $G$ and min $st$-cuts in the dual $G^*$ of a plane graph: let $f$ and $g$ be distinct faces in $G$ and let $f^*$ and $g^*$ be the corresponding dual vertices in $G^*$, respectively. Let $C$ be a shortest clockwise cycle in $G$ such that $f$ is outside of $C$ and $g$ is inside. Then there is a min $f^*g^*$-cut in $G^*$ such that the arcs crossing this cut are exactly the (dual) arcs of $C$. It is now easy to see that the dual of the graph in Figure 1 has $\Theta(n^2)$ distinct min cuts. Moreover, the total size of the min cut-sets in this example is $\Theta(n^3)$.

At first sight it might thus seem impossible to compute the capacities of all min cuts in $o(n^3)$ time. Indeed, repeated applications of the single-source single-sink max flow algorithm of Borradaile and Klein [3] leads to a time bound of $O(n^3 \log n)$. Arikati et al. [1] computed all of these capacities in $O(n^2 + \gamma^3 \log \gamma)$ time where $\gamma$ is the minimum number of *hammocks*, a special kind of outerplanar subgraphs, into which $G$ can be partitioned. However, $\gamma$ might be $\Theta(n)$ and the running time of this algorithm is also $O(n^3 \log n)$ in this case. Nevertheless, we are able to significantly improve this to $O(n^2 \log^3 n)$ which is optimal up to logarithmic factors. In fact, we show something stronger: for fixed source $s$, max $st$-flow values for all $n - 1$ sinks $t \neq s$ can be found in a total of $O(n \log^3 n)$ time.

Our algorithm can be changed to a data structure that can report the arcs crossing the min cuts, i.e., min cut-sets. Given a fixed source $s$ it requires $O(n^{3/2} \log^{3/2} n)$ preprocessing time. Afterwards, it can report a min cut-set $C$ for $s$ and any sink $t \neq s$ in $O(|C| \log \log n)$ time. Hence, the total time to compute all min cut-sets $\mathcal{C}$ in the planar digraph is $O(n^{5/2} \log^{3/2} n + \sum_{C \in \mathcal{C}} |C| \log \log n)$. With a longer preprocessing of $O(n^{5/3} \log^{2/3} n)$ time, we can decrease the report time to $O(|C|)$.

Our result shows that in planar digraphs all $O(n^2)$ max flow values can be computed in time essentially equal to the time for finding a linear number of max flows. This joins the result of Arikati et al. [1] that showed an $O(n^2)$ algorithm for the same problem for *bounded treewidth* digraphs. For general $n$-vertex digraphs, Hao and Orlin [18] showed how to compute $O(n)$ max flows in the time it takes for a single max flow computation and used this result to find a (global) min cut. Based on this and our result for planar digraphs, we conjecture that also in general digraphs all $O(n^2)$ max $st$-flow values can be computed faster than the time required for computing $O(n^2)$ max $st$-flows separately. The result of Hao and Orlin does not resolve this conjecture as the $O(n)$ source/sink pairs considered by their algorithm are in a sense arbitrary; in particular, their result cannot be used to efficiently find max $st$-flow values for fixed $s$ and all $t \neq s$.

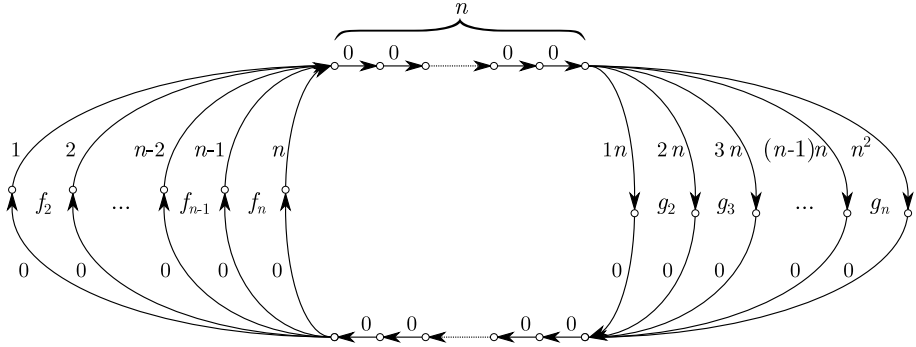

Figure 1: Example of a planar digraph on $4n$ vertices for which min $f_i^* g_j^*$-cuts in the dual are pairwise distinct, $2 \leq i, j \leq n$. There are thus $\Theta(n^2)$ min cuts and each has $\Theta(n)$ arcs for a total of $\Theta(n^3)$. In order to make the dual graph simple one can triangulate the above graph with infinite weight edges.

When all arc capacities are equal to a single unit, we get the *arc connectivity* problem. For this special case of our problem, Cheung et al. [9] showed an $O(m^\omega)$ all-pairs arc connectivity algorithm, for an $m$-arc digraph, where $\omega$ is the matrix multiplication exponent. This is faster than running the $O(\min\{m^{1/2}, n^{2/3}\} \cdot m)$ algorithm for $st$-arc connectivity of Even and Tarjan [11] for every pair of vertices when $m = O(n^{1.94})$, using the currently best known value of $\omega < 2.3727$ [28]. Cheung et al. also showed an $O(d^{\omega-2} n^{\omega/2+1})$ algorithm for the same problem for *well-separable* digraphs, which also include planar graphs, with maximum degree $d$.

Cabello et al. [7] gave a simple algorithm for computing the *shortest non-contractible cycle* in a directed graph embedded in a sphere with $b$ boundaries, which uses $b$ minimum cut computations with a fixed source in the dual planar graph. Our result improves the time bound of this algorithm from $O(bn \log n)$ to $O(n \log^3 n)$ when $b = \omega(\log^2 n)$. In a similar way, our algorithm can be used to find the *shortest clockwise cycle* in an embedded directed planar graph, by finding the min $st$-cut in the dual planar graph where $s$ is the vertex dual to the inifinite face of the original graph.

The organization of the paper is as follows. In Section 2, we make some definitions and simplifying assumptions. In Section 3, the overview of the algorithm for our problem is presented. Our algorithm is guided by a recursive decomposition of the input graph. It recursively computes max preflows to the outer boundaries of all subgraphs in the decomposition using previously computed max preflows to outer boundaries of the parent subgraphs. When the algorithm reaches a leaf of the recursive decomposition, a max preflow to a sink $t$ contained in that leaf is found. The value of a max $st$-flow can then be identified as the amount of preflow into $t$. A straightforward implementation of this algorithm will not lead to a near-linear time bound. In Section 4, we show how to efficiently implement various steps of the algorithm. An important tool here is a modification of a flow fixing procedure from [4]. In Section 5, we show how the min $st$-cut-sets themselves can be reported. Finally, in Section 6 we give some conclusions and suggest future directions of research.

# 2 Preliminaries

Let $G = (V, E)$ be a simple planar digraph, where $V$ is the vertex set and $E \subseteq V \times V$ is the set of arcs. Let $n = |V|$. Since $G$ is planar, we have $|E| = O(n)$. We assume that for every arc $e = (u, v) \in E$ the *reversed arc* $rev(e) = (v, u)$ is also in $E$. This assumption can be easily satisfied for planar graphs, by using the same embedding for $e$ and $rev(e)$. By *edge* we mean a pair of two opposite arcs, $e$ and $rev(e)$, which share their embedding in the plane.

We use the standard definition of the *dual planar graph* $G^*$, that is, every dual vertex corresponds to a primal face, and two dual vertices are connected by an edge if and only if there is an edge between the two corresponding primal faces. We orient the dual arcs such that the dual arc of a primal arc $e$ is oriented from the right side of $e$ to its left side. If the primal is a flow network with capacities on arcs, the dual arcs have weights equal to these capacities.

## 2.1 Flows

For graph $G$ the capacities of the arcs are given by a *capacity function* $c : E \to \mathbb{R}^+$. Let $s \in V$ be a source and let $t \in V \setminus \{s\}$ be a sink. We define a *flow assignment* to be a function $f : E \to \mathbb{R}$ satisfying *antisymmetry*, i.e., for all $e \in E$ we require $f(e) = -f(rev(e))$. A flow assignment is called an *st-pseudoflow* if for all arcs $e \in E$ we have $f(e) \leq c(e)$. The inflow of an $st$-pseudoflow $f$ for a vertex $v \in V$ is defined as:

$$\text{inflow}_f(v) = \sum_{(u,v) \in E} f(u, v),$$

We say that $st$-pseudoflow $f$ has *excess* in $v$ when $\text{inflow}_f(v) > 0$ and in this case we refer to $\text{inflow}_f(v)$ as its *excess (value)*. Similarly, we say that $f$ has *deficit* in $v$ when $\text{inflow}_f(v) < 0$ and $\text{inflow}_f(v)$ is its *deficit (value)*. An *st-preflow* is an $st$-pseudoflow such that no vertex in $V \setminus \{s\}$ has deficit. An *st-flow* is an $st$-preflow such that no vertex in $V \setminus \{t\}$ has excess. When no confusion arises, we shall omit $s$ and $t$ and simply talk about pseudoflows, preflows, and flows. A *circulation* is a pseudoflow such that for every vertex $v \in V$, $\text{inflow}_f(v) = 0$. A circulation does not have a source or a sink.

The *value* of a preflow $f$ is given as $\text{inflow}_f(t)$. Finally, a *max pseudoflow (preflow)* is a pseudoflow (preflow) such that there is no residual path that starts at a source or at a vertex with excess and ends at a sink or at a vertex with deficit. Observe that the value of a max preflow equals the value of a max flow. We can *limit* the value of a max flow by some value $d$, by adding a new vertex $s'$ and an arc $(s', s)$ of capacity $d$ and regarding $s'$ as a source instead of $s$.

An *st-cut* is a partition of the vertex set $V$ into two sets $S$ and $V \setminus S$, such that $s \in S$ and $t \in V \setminus S$. The set of arcs going from $S$ to $V \setminus S$ is called the *cut-set*. The *value* of an $st$-cut is the sum of capacities of all arcs in the cut-set. The max-flow min-cut theorem [13] states that the value of the max $st$-flow is equal to the value of the min $st$-cut.

Let $M$ be some fixed value greater than the sum of all capacities in the graph, that is $M > \sum_{e \in E} c(e)$. When we say that we assign *infinite capacity* to an arc, for example for vertex splitting, we in fact assign finite capacity of value $M$. This way we keep the flow assignment finite.

## 2.2 Pieces and Recursive Decomposition

Let $G$ be a plane graph. We assume, for simplicity, that $G$ is triangulated and has bounded degree. Both can be satisfied simultaneously using a standard vertex-splitting argument to get degree three and then triangulating each face such that each vertex degree in that face is increased by at most 2. In Section 5, where we report the cut-sets in time proportional to their sizes, we report the cut-set from the original graph without these simplification assumptions.

A *piece* is a subgraph of $G$ with the same embedding as $G$. A vertex of $P$ is called a *boundary vertex* of $P$ if it is incident in $G$ to a vertex not belonging to $P$. All other vertices of $P$ are called *interior vertices* of $P$. We let $\partial P$ denote the set of boundary vertices of $P$. A *hole* of $P$ is a face of $P$ which is not a face of $G$. In certain places, we shall regard a hole as the subgraph of $G$ embedded inside of it. It should be clear from context what is meant.

A *decomposition* of a piece $P$ is a set of sub-pieces $P_1, \ldots, P_k$ such that the union of the vertex sets of these sub-pieces is the vertex set of $P$ and such that every edge of $P$ is contained in a unique sub-piece. We define every boundary vertex of $P$ to be a boundary vertex of every sub-piece $P_i$ that contains it. We change the standard definition of a decomposition a little and allow two sub-pieces to share edges. Edges that connect boundary vertices of a sub-piece are always included into this sub-piece, even if these edges belong also to another sub-piece. This does not increase the size complexity of the decomposition, and will simplify the discussion of the dual graph of the sub-piece.

A *recursive decomposition* of $G$ is obtained by first identifying a decomposition of $G$ and then recursing on each sub-piece. The recursion stops when pieces of constant size are obtained. The recursive decomposition of $G$ is the collection of pieces constructed over all levels of the recursion. We are interested in a special type of recursive decomposition satisfying the following: a piece $P$ with $p$ vertices and $b$ boundary vertices is divided into a constant number of connected sub-pieces each containing at most $\frac{1}{2}p$ vertices, at most $\frac{1}{2}b$ boundary vertices inherited from $P$, and at most $O(\sqrt{p})$ additional boundary vertices. In addition, we require that each piece has only a constant number of holes. When we refer to a recursive decomposition of $G$ in the following, we shall assume that it is of this special type. Parent/child and ancestor/descendant relationships between pieces correspond to their relationships in the decomposition tree.

Obtaining a recursive decomposition that satisfies all of the above conditions is non-trivial. The assumptions that subpieces contain at most $\frac{1}{2}p$ vertices, at most $\frac{1}{2}b$ boundary vertices inherited from $P$, and at most $O(\sqrt{p})$ additional boundary vertices can be satisfied by finding an $r$-division of $P$, for $r = \frac{1}{2}p$; see [21] for details. A construction of connected pieces, as we require, is given in [5]. In order to ensure that pieces are connected, we allow $O(n)$ new edges to be added while maintaining planarity and the chosen embedding of $G$. The arcs corresponding to these edges have zero capacity and thus do not affect max flows or min cuts.

The results in the remainder of this subsection are taken from [5].

**Lemma 1.** *A recursive decomposition as described above can be computed in $O(n \log n)$ time.*

In order to bound the running time of our algorithms we need the following lemma.

**Lemma 2.** *Let $\mathcal{P}$ be the set of pieces in a recursive decomposition of $G$. Then $\sum_{P \in \mathcal{P}} |P| = O(n \log n)$ and $\sum_{P \in \mathcal{P}} |\partial P|^2 = O(n \log n)$.*

The proof of Lemma 2 follows by combining the following two lemmas with the fact that the depth of the recursive decomposition is $O(\log n)$.

**Lemma 3.** *Let $\mathcal{P}_i$ be the set of pieces in level $i$ of a recursive decomposition of $G$. Then $\sum_{P \in \mathcal{P}_i} |P| = O(n)$.*

*Proof.* Let $c_{\min}$ be the constant such that pieces of size at most $c_{\min}$ are not decomposed further in a recursive decomposition. We may assume that no piece has size less than $\frac{1}{2}c_{\min}$. Let $L(p)$ denote the total number of vertices (counted with multiplicity) in the leaf-pieces of the recursive decomposition of an $p$-vertex piece $P$. We will show that $L(p) \leq c_1 p - c_2 \sqrt{p}$ for suitable constants $c_1$ and $c_2$. The lemma will follow since the number of vertices (counting multiplicity) in any level of the recursive decomposition is dominated by the number of vertices in the leaves; this is bounded by $L(n)$ which is $O(n)$.

We prove that $L(p) \leq c_1 p - c_2 \sqrt{p}$ by induction. In the base cases, in which a piece of size $p$ is a leaf of the recursive decomposition (so $p \in [\frac{1}{2}c_{\min}, c_{\min}]$), $L(p) = p$. This is bounded by $c_1 p - c_2 \sqrt{p}$ so long as $c_{\min} \leq c_1(\frac{1}{2}c_{\min}) - c_2\sqrt{c_{\min}}$. Setting $c_{\min} \geq \left(c_2/(\frac{1}{2}c_1 - 1)\right)^2$ guarantees this.

Now consider a piece $P$ of size $p > c_{\min}$. Assume inductively that the claim holds for all values smaller than $p$. Let $p_i$ be the size of the $i^{th}$ child of $P$; $P$ has $N = O(1)$ children. By the inductive hypothesis we get:

$$L(p) = \sum_i L(p_i) \leq \sum_i (c_1 p_i - c_2 \sqrt{p_i}) = c_1 \sum_i p_i - c_2 \sum_i \sqrt{p_i} \tag{1}$$

We lower bound $\sum_i \sqrt{p_i}$ by observing that $\sum_i \sqrt{p_i}$ can only be as small as allowed by $p \leq \sum_i p_i$, $p_i \in [0, \frac{p}{2}]$. The minimum value occurs when two of the $p_i$'s are equal to $\frac{p}{2}$ and all others are zero, giving:

$$\sum_i \sqrt{p_i} \geq 2\sqrt{\frac{p}{2}} = \sqrt{2p} \tag{2}$$

Also note that when $P$ is decomposed, only the new boundary vertices are replicated among the child pieces. Let us denote by $\sigma$ the constant of the planar separator theorem, then

$$\sum_i |p_i| \leq |p| + \sigma N \sqrt{|p|}. \tag{3}$$

Combining Equations (3), (1), and (2), we get that

$$L(p) \leq c_1(p + \sigma N \sqrt{p}) - c_2 \sqrt{2p} = c_1 p - c_2 \left(\sqrt{2} - \frac{c_1}{c_2}\sigma N\right) \sqrt{p}$$

This completes the induction for $c_2$ sufficiently larger than $c_1 \sigma N$. □

**Lemma 4.** *Let $\mathcal{P}_i$ be the set of pieces in level $i$ of a recursive decomposition of $G$. Then $\sum_{P \in \mathcal{P}_i} |\partial P|^2 = O(n)$.*

*Proof.* Let $P$ be any piece or the whole graph. By adding dummy boundary vertices that never become boundary nodes in the children, we may assume that:

$$|\partial P| \geq c\sqrt{|P|} \tag{4}$$

where $c$ is a constant that we will pick below. In the case when $P$ is the whole graph $G$ we write $\partial G$ do refer to these dummy vertices. We will show that for any piece $P$ with children $P_1, \ldots, P_N$:

$$\sum_j |\partial P_j|^2 \leq |\partial P|^2. \tag{5}$$

The lemma follows from this because, by summing over all pieces in a level we get,

$$\sum_{P \in \mathcal{P}_i} |\partial P|^2 \leq \sum_{P \in \mathcal{P}_{i-1}} |\partial P|^2 \leq \cdots \leq \sum_{P \in \mathcal{P}_0} |\partial P|^2 = |\partial G|^2$$

Since $G$ has only dummy boundary vertices, $|\partial G|^2 = c(\sqrt{|G|})^2$, which is $O(n)$, as desired.

We now prove Equation (5). In the next equation, the first and second inequalities follow from the definition of recursive decomposition and Equation (4), respectively:

$$|\partial P_j| \leq \frac{1}{2}|\partial P| + \sigma\sqrt{|P|} \leq \left(\frac{1}{2} + \frac{1}{c}\right)|\partial P| \qquad (6)$$

Similarly, as in previous lemma we observe that only the new boundary vertices are shared between the children, so:

$$\sum_i |\partial P_i| \leq |\partial P| + \sigma N\sqrt{|P|}. \qquad (7)$$

Combining it all together we get:

$$\begin{aligned}
\sum_j |\partial P_j|^2 &\leq \left(\frac{1}{2} + \frac{1}{c}\right)|\partial P|\sum_j |\partial P_j| &\text{by Equation (6)} \\
&\leq \left(\frac{1}{2} + \frac{1}{c}\right)|\partial P|\left(|\partial P| + \sigma N\sqrt{|P|}\right) &\text{by Equation (7)} \\
&\leq \left(\frac{1}{2} + \frac{1}{c}\right)|\partial P|\left(|\partial P| + \sigma N\frac{|\partial P|}{c}\right) &\text{by Equation (4)} \\
&= \left(\frac{1}{2} + \frac{1}{c}\right)\left(1 + \frac{\sigma N}{c}\right)|\partial P|^2 \\
&\leq |\partial P|^2 &\text{for constant } c \text{ sufficiently large.}
\end{aligned}$$

This completes the proof. □

### 2.3 Nesting of Outer Boundaries

Let $\mathcal{P}$ be the set of pieces in the recursive decomposition. We may assume that the fixed source $s$ is not a boundary vertex of any piece of $\mathcal{P}$. We can always add a new vertex $s'$ to $G$ with an infinite capacity edge to $s$ and regard $s'$ instead of $s$. This guarantees that the following definitions are unambiguous. If a hole $H$ of $P$ contains $s$, we refer to $H$ as the *outside* of $P$. The subgraph of $G$ contained in the outside of $P$ is denoted $\text{ext}(P)$. The *outer boundary* of $P$ is the set of boundary vertices of $P$ contained in the outside of $P$. Note that the outer boundary of a piece is contained in a single cycle of that piece.

For simplicity, we also assume that each sink belongs to a piece having an outside. This is true for every vertex that belongs to a piece that $s$ does not belong to, that is for all vertices of the graph except for a constant number of vertices which belong to the same leaf piece of the recursive decomposition as $s$. For this set of sinks we can compute max $st$-flow for every sink in $O(n \log n)$ total time using the algorithm of Borradaile and Klein [3].

These assumptions guarantee that the outer boundaries of the recursive decomposition has a rather simple structure as illustrated in Figure 2(a). The outer boundaries nest and form a laminar family, i.e., a forest. Let the nesting of outer boundaries be represented by a forest $\mathcal{F}$. The outer boundary $C$ is an ancestor in $\mathcal{F}$ of an outer boundary $C'$ if the $s$-side of $C'$ contains the $s$-side of $C$. Each outer boundary $C$ is represented in the forest only once, even if it is shared by multiple pieces. For an example, see Figure 2(b).

For any $C \in \mathcal{F}$ which is not a root in $\mathcal{F}$, we define the *outer piece* of $C$ as follows:

1. If $C$ bounds a hole $H$ of a piece $P \in \mathcal{P}$, which is not the outside of $P$, then the deepest piece $P$ in the recursive decomposition which satisfies this condition is the outer piece of $C$ (see piece $a$ in Figure 3).

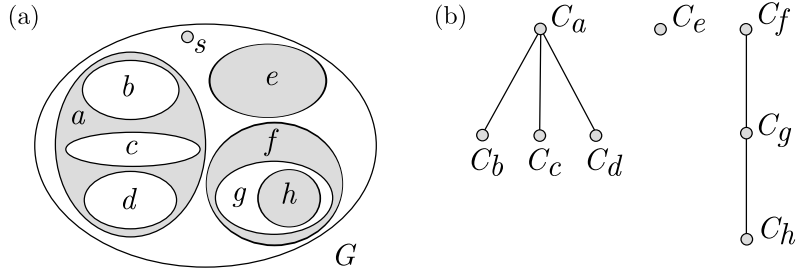

Figure 2: Figure (a) shows a recursive decomposition that satisfies out all assumptions. In Figure (b), the corresponding forest $\mathcal{F}$ is depicted.

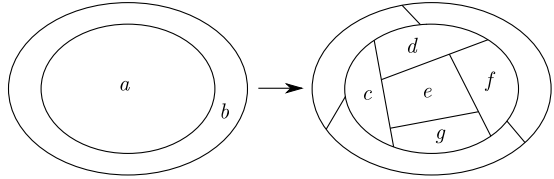

Figure 3: A single step of building the recursive decomposition. The piece is divided by finding an $r$-division. The outer piece of the outer boundary of $a$ is $b$. The outer piece of the outer boundary of $c$ is the part of $a$ outside of $c$, which contains $d$, $e$, $f$, $g$, and the outer boundary of $a$.

2. Otherwise, $C$ does not bound a hole of any piece, except the outside of the pieces of which it is the outer boundary. Let $P$ be a piece such that $C$ is the outer boundary of $P$. Let $C'$ be the parent of $C$ in $\mathcal{F}$ and let $P'$ be the deepest piece in the recursive decomposition such that $C'$ is the outer boundary of $P'$. We define the outer piece of $C$ to be the intersection of $P'$ with $\text{ext}(P)$ (see piece $c$ in Figure 3). Note that in this case the outer piece of $C$ is not a piece of the recursive decomposition.

To simplify our description, we shall assume in the following that all cycles $C \in \mathcal{F}$ have outer pieces of the first type. In particular, we assume that all outer pieces belong to the recursive decomposition. Dealing with outer pieces of the second type does not increase our time bound. We simply can include these additional outer pieces into $\mathcal{P}$. The following lemma implies that handling such pieces does not affect the asymptotic running time of the algorithm.

**Lemma 5.** *The number of trees in $|\mathcal{F}|$ is $O(1)$ and each tree in $\mathcal{F}$ has constant degree.*

*Proof.* By assumption, $s$ is not a boundary vertex of any piece. For any outer boundary $C$ corresponding to a root in $\mathcal{F}$, there is no other outer boundary in $\mathcal{F}$ separating $C$ and $s$. Hence, every such $C$ must lie inside a hole in the same piece $P$. Since $P$ has only $O(1)$ holes, the first part of the lemma follows.

For a node of $\mathcal{F}$ to be a child of another node in $\mathcal{F}$, the outer boundaries corresponding to these nodes must share vertices with holes of the same piece. The second part of the lemma follows, again since a piece only has $O(1)$ holes and each piece is divided into a constant number of connected subpieces. □

## 2.4 Dense Distance Dual Graphs

For a piece $P$, Fakcharoenphol and Rao [12] define the *dense distance graph* of $P$ as the complete directed graph on the set of boundary vertices of $P$, representing shortest path distances among them in $P$. We apply a similar construction to the dual graph, guided by the recursive decomposition. We define dense distance dual graphs for the pieces of the decomposition and for the holes of these pieces. We define a *boundary face* (of a piece or of a hole) to be a face incident to a boundary vertex (of the same piece or of the same hole). Our definition of dense distance dual graph is over the set of dual vertices corresponding to boundary faces.

Let $P$ be a piece in the recursive decomposition, denote by $G^*[P^*]$ the subgraph of $G^*$ induced by the dual vertices corresponding to the faces of $P$. The *internal dense distance dual graph* IDDG$^*(P)$ of $P$ is the complete directed graph over the set of dual vertices corresponding to boundary faces of $P$. An arc $(f_1, f_2)$ in IDDG$^*(P)$ has weight equal to the (possibly infinite) shortest path distance from $f_1$ to $f_2$ in $G^*[P^*]$. Observe that due to our assumption that vertices have constant degree, the number of vertices in IDDG$^*(P)$ is $O(|\partial P|)$. The number of edges in IDDG$^*(P)$ is $O(|\partial P|^2)$, and thus by Lemma 2, the total size of the internal dense distance dual graphs of all pieces of the recursive decomposition of $G$ is $O(n \log n)$.

Let $H$ be a hole of $P$. We define the *dense distance dual graph* of $H$, denoted DDG$^*(H)$, in a similar way. It is a complete directed graph on the set of dual vertices corresponding to boundary faces of $H$. Each arc $(f_1, f_2)$ of DDG$^*(H)$ has weight equal to the shortest path distance in from dual vertex $f_1$ to the dual vertex $f_2$ in the subgraph of $G^*$ induced by the faces of $H$. Lemma 2 and our assumption that vertices have constant degree imply that the total number of edges of DDG$^*(H)$ over all holes of a single piece $P$ is $O(|\partial P|^2)$ and that the total number of edges over all dense distance dual graphs of all holes in the recursive decomposition of $G$ is $O(n \log n)$. We do not compute DDG$^*(H)$ for the hole $H$ of $P$ such that $H = \text{ext}(P)$.

The advantage of Fakcharoenphol and Rao's dense distance graphs is a faster implementation of Dijkstra's algorithm which we refer to as *FR-Dijkstra*. This algorithm can find a shortest path tree of a given vertex in a graph composed of dense distance graphs of pieces of a planar graph and additional arcs from the (original) planar graph in $O(b \log^2 b)$ time, where $b$ is the sum of the total number of vertices in the dense distance graphs, counted with multiplicity, and the total number of additional original arcs. We can use the implementation of FR-Dijkstra without any change also on dense distance dual graphs together with dual arcs from $G^*$.

Following Fakcharoenphol and Rao, we use FR-Dijkstra also for the construction of the dense distance dual graphs. Klein [27] gave a faster construction of dense distance graphs that construct dense distance graph in $O(|P| \log |P|)$ time for a piece $P$ with $\partial P = O(\sqrt{P})$, however we cannot use it for computing dense distance dual graphs of holes, since they need not obey this connection between the number of boundary faces and the size of the hole. We give more details about the algorithm of [27] below.

**Lemma 6.** *We can compute for all pieces $P \in \mathcal{P}$ the graph* IDDG$^*(P)$ *and the graphs* DDG$^*(H)$ *for every hole $H$ of $P$ other than* $\text{ext}(P)$, *in total of $O(n \log^3 n)$ time.*

*Proof.* First we compute IDDG$^*(P)$ for every $P \in \mathcal{P}$ using Klein's construction [27] in $O(n \log^2 n)$ total time. We proceed bottom-up on each tree $T \in \mathcal{F}$. Note that all the holes that we are interested in are bounded by some outer boundary in $\mathcal{F}$. For each $C \in T$, we consider the hole $H$ inside $C$ and compute DDG$^*(H)$ as follows. If $C$ is a leaf of $T$ then the construction is trivial, since the size of $H$ is bounded by a constant. Otherwise, let $P_C$ be a piece whose outer boundary is $C$. We build a graph $D^*$ composed of IDDG$^*(P_C)$ and dense dual distance graphs DDG$^*(H')$ for all holes $H' \neq \text{ext}(P_C)$ of $P_C$. Note that for every such $H'$, the outer boundary that bounds $H'$ is inside the outer boundary $C$ which bounds $H$. Thus we have already computed DDG$^*(H')$. We compute DDG$^*(H)$ by running FR-Dijkstra in $D^*$ from each of the $O(|C|) = O(|\partial P_C|)$ boundary faces of $H$. Each run takes $O(|\partial P_C| \log^2 |\partial P_C|)$ time, so in total we require $O(|\partial P_C|^2 \log^2 |\partial P_C|)$ time for each outer boundary $C$. Applying Lemma 2 for the whole recursion we get $O(n \log^3 n)$ total time. □

We will use FR-Dijkstra also for the rest of our algorithm. Using FR-Dijkstra with *reduced*

*lengths* defined by a *potential function* (see Section 4.1) will turn out to be very useful for our purposes. The original implementation of FR-Dijkstra by Fakcharoenphol and Rao [12] does not support reduced lengths. Therefore, we use an extension of FR-Dijkstra by Kaplan et al. [25] which allows using them.

## 2.5 Multiple Source Shortest Path

Another useful tool which we use in our paper is the multiple source shortest path (MSSP) data structure of Klein [27], which we already mentioned above. The MSSP data structure computes efficiently a set of distances such that all sources are different vertices along a single face of a planar graph. The data structure computes $k$ distances in $O((n+k)\log n)$ time. In addition, it can also produce an $O(n)$ space data structure, which reports the shortest paths themselves in $O(\ell \log \log n)$ time per path of length $\ell$.

We use the MSSP data structure to compute dense distance dual graphs and for reporting, in Section 5, the dual paths that constitute parts of the cut-sets which we report. In a typical use of the MSSP data structure by our algorithm, we compute distances among boundary faces adjacent to a boundary cycle $C$ of a piece $P$ that bounds a hole $H_C$, inside the hole (we also use the MSSP for distances inside $P$; this case is similar). The internal part of the piece $P$ is not part of $H_C$, and there is a single face in the embedding of $H_C$ that corresponds to $P$. In the dual graph of $H_C$, all the dual vertices that we are interested in, which correspond to the boundary faces, are adjacent to the single vertex that represents the face of $H_C$ that corresponds to $P$. Once we remove this vertex, together with the edges dual to $C$, which are all the edges adjacent to this vertex, all the vertices that are dual to the boundary faces are on a single dual face and we can use the MSSP algorithm on them.

## 3 Single Source - All Sinks Max Flow

For a fixed source $s$, we wish to find the value of a max $st$-flow in $G$ for every sink $t \in V \setminus \{s\}$. In this section we describe an algorithm which is based on computing the value of a max flow from $s$ to every outer boundary in $\mathcal{F}$ in a recursive fashion. From this we can easily find the values we are looking for. In Section 4, we describe an efficient implementation of the algorithm.

### 3.1 Max Preflow in a Separated Graph

First, we consider a more specific problem, which enables us to implement the recursive step of our algorithm. Let $G_1 = (V_1, E_1)$ and $G_2 = (V_2, E_2)$ be non-empty subgraphs of $G$, where $V_1 \cup V_2 = V$, $E_1 \cup E_2 = E$ and $E_1 \cap E_2$ is a simple cycle $C$ with vertex set $V_1 \cap V_2$. We shall identify $C$ with $V_1 \cap V_2$. Assume that $s \in V_1 \setminus C$ and $t \in V_2 \setminus C$. This assumption can be made without loss of generality. Algorithm MAXPREFLOWINSEPARATEDGRAPH finds a max $st$-preflow in $G$, given a max $sC$-preflow $f_{sC}$ in $G_1$:

Algorithm MAXPREFLOWINSEPARATEDGRAPH($G_1, G_2, f_{sC}$):

1. Find a max $Ct$-flow $f_{Ct}$ in $G_2$. Denote the value of this flow by $d_{Ct}$.

2. Add the two flow assignments $f_{sC}$ and $f_{Ct}$ to form a pseudoflow in $G$.

3. Send flow among vertices of $C$ until there is no residual path from a vertex $u \in C$ with excess to a vertex $v \in C$ with deficit; update the residual network accordingly.

4. If there are vertices in $C$ with remaining deficit, let $-\ell$ be the sum of deficits over all vertices of $C$ with a deficit; reset the flow on $G_2$, and rerun the algorithm with a limit of $d_{Ct} - \ell$ on the $Ct$-flow pushed in step 1.

The above algorithm is based on the *flow partition* scheme of Johnson and Venkatesan [24], which was also used in [4]. Our implementation is a little different from the original scheme in its last step. First, we do not return excess flow from $C$ to $s$; we are not required to do so, since we are looking for a preflow. Second, we do not return deficit flow from $C$ to $t$. Instead we recompute a $Ct$-flow with value limited to $d_{Ct} - \ell$. This computation is equivalent, but as we show below, we can implement it faster.

Notice that since edges of $E_1 \cap E_2$ are incident only to vertices of $C$, we may assume that both $f_{sC}$ and $f_{Ct}$ do not assign flow to any of these edges, since the vertices of $C$ are all sinks of the first and sources of the second. Therefore at step 2 we indeed obtain a pseudoflow that does not violate the arc capacities.

We shall refer to an algorithm implementing step 3, that is sending as much flow as possible from excess vertices of $C$ to deficit vertices of $C$, as a *flow fixing procedure* along $C$ in the residual network.

## 3.2 Max Preflow in Pieces

The main algorithm of this section, which we call Algorithm MAXPREFLOWTOBOUNDARIES, uses Algorithm MAXPREFLOWINSEPARATEDGRAPH in recursion guided by the nesting of outer boundaries. The problem that Algorithm MAXPREFLOWTOBOUNDARIES solves is to identify, for each piece $P$, a max preflow from $s$ to the outer boundary of $P$. This max preflow is entirely outside of $P$.

Instead of considering every piece, it is enough to consider only the unique outer boundaries in $T$, since if two pieces share the same outer boundary, we do not need to compute a max preflow from $s$ to this outer boundary twice. Our algorithm traverses each tree $T \in \mathcal{F}$ from root to leaves. Let $C$ be the outer boundary corresponding to the root node of $T$. The algorithm gets the max preflow $f_{sC}$ from $s$ to $C$ outside of $C$ as a parameter. This max preflow can be computed by embedding a super-sink $t'$ inside $C$, connecting every vertex of $C$ to $t'$ by infinite capacity arcs, applying a max $st'$-flow algorithm for planar graphs, and removing $t'$.

Algorithm MAXPREFLOWTOBOUNDARIES($T$,$C$,$f_{sC}$):

1. For every child $C'$ of $C$ in $T$, perform the following steps:

2. Add a super-sink $t'$ inside $C'$ connecting every vertex of $C'$ to $t'$ using infinite capacity arcs.

3. Apply Algorithm MAXPREFLOWINSEPARATEDGRAPH($G_1, G_2, f_{sC}$) to find a max preflow $f_{st'}$ from $s$ to $t'$, where $G_1$ is the part of the graph outside of $C$ and $G_2$ is the part inside $C$ and outside $C'$.

4. Remove $t'$ and the arcs incident to it, from $f_{st'}$ to obtain $f_{sC'}$, a max $sC'$-preflow.

5. Apply MAXPREFLOWTOBOUNDARIES($T$,$C'$,$f_{sC'}$).

Algorithm MAXPREFLOWTOBOUNDARIES considers every outer boundary $C'$ which is a child of $C$ in $T$. The max preflow from $s$ to $C$ outside of $C$ is given as a parameter of the procedure. The algorithm finds a max preflow from $s$ to $C'$, using a super-sink $t'$ inside $C'$ and infinite capacity arcs as before. This way the problem becomes finding a max preflow from $s$ to $t'$. (Note that $C$ and $C'$ might share some vertices, in this case we do not connect the shared vertices to $t'$, to avoid flow of infinite value.) We find the max preflow from $s$ to $t'$ using Algorithm MAXPREFLOWINSEPARATEDGRAPH. We can regard $C$ as a separator in the graph, where $G_1$ is the $s$-side (outside) of $C$ and $G_2$ is the $t$-side (where $C'$ is). In the rest of this section and in the following section we focus on the fast implementation of the above steps.

Algorithm MaxPreflowToBoundaries computes a max preflow from $s$ to the outer boundary of $P$, in $\text{ext}(P)$, for every piece $P$ in the recursive decomposition. Every sink $t$ is contained in some piece of constant size $P_t$. We find a max preflow from $s$ to each sink $t$, using algorithm MaxPreflowInSeparatedGraph with the max flow $f_{sC}$ as a parameter, where $C$ is the outer boundary of $P_t$. From this, we can obtain the value of a max $st$-flow for each sink $t \neq s$ in $G$, as required, since this value is the same as the value of the max $st$-preflow that we found.

### 3.3 Pushing Flow Through Holes

Let $C$, $C'$, and $G' = G_1 + G_2$ be as defined in the description of algorithm MaxPreflowToBoundaries. Let $P$ be the outer piece of $C'$. Note that $C$ is the outer boundary of $P$. In step 1 of Algorithm MaxPreflowInSeparatedGraph we push flow from $C$ to $C'$ in $G_2$. A simple way to implement this step is using the max flow algorithm of Borradaile and Klein [3] in $G_2$ with a super-source connected to $C$ and a super-sink connected to $C' \setminus C$, in $O(|G_2|\log|G_2|)$ time. If $P$ has no holes except of the holes bounded by $C$ and by $C'$, then $G_2$ is $P$, and we can use the simple solution. However, if there are more holes in $P$, other than these two, then $G_2$ might be much bigger than $P$, since it should contain also the parts of the graph inside the additional holes. We cannot afford to spend that much time in this step. Rather, we would like to bound the amount of work here by the size of $P$, so that we can use Lemma 2 to bound the total time of our algorithm.

Let $H_1, \ldots, H_k$ denote the holes of $P$ bounded by neither $C$ nor $C'$ and let $C_1, \ldots, C_k$ be the boundaries of these holes. We observe that $G_2 = P + H_1 + \cdots + H_k$. In this subsection, we show how to decompose the computation of the max flow found in step 1 of Algorithm MaxPreflowInSeparatedGraph into smaller steps according to this partition of $G_2$, such that these steps consist of max flow computations inside $P$ and of calls to the flow fixing procedure. In Section 4, we will give an efficient implementation for these steps.

**The single additional hole case**  We first consider the case where $k = 1$. Let $H = H_1$ and let $C_H = C_1$ be the boundary of $H$. We give here an algorithm PushFlowThroughHoles($P, C, C', H$) which finds a max flow inside $P + H$ from $C$ to $C'$. The algorithm is as follows (each step uses the residual network for the flow found in earlier steps):

Algorithm PushFlowThroughHoles($P, C, C', H$):

1. Compute a max flow in $P$ from $C$ to $C'$.

2. Compute a max flow in $P$ from $C$ to $C_H$. Denote the value of this flow by $d_2$.

3. Compute a max flow in $P$ from $C_H$ to $C'$. Denote the value of this flow by $d_3$.

4. Run a flow fixing procedure along $C_H$ in $P + H$.

5. Let $\ell_2$ be the sum of excesses over vertices of $C_H$ with an excess, and let $-\ell_3$ be the sum of deficits over all vertices of $C_H$ with a deficit; reset the flow to zero in $P + H$, and rerun the previous steps with a limits of $d_2 - \ell_2$ and $d_3 - \ell_3$ on the flow pushed in steps 2 and 3, respectively.

**Lemma 7.** *Algorithm* PushFlowThroughHoles($P$, $C$, $C'$, $H$) *finds a max flow from $C$ to $C'$ in $P + H$.*

*Proof.* We will show that after (the first execution of) step 4, we have a pseudoflow with no residual paths in $P + H$ from $C$ to $C'$, from $C$ to deficit vertices on $C_H$, from excess vertices on $C_H$ to $C'$, or from excess vertices on $C_H$ to deficit vertices on $C_H$. This suffices to show that at termination we have a max flow from $C$ to $C'$ in $P + H$, since step 5 in fact only returns excess to $s$ and deficit to $t$, as in the flow partition scheme of Johnson and Venkatesan [24].

Let us identify a cut with the corresponding cut-set, that is the set of arcs crossing the cut. After step 1, there is a saturated cut $K_1$ in $P$ separating $C$ from $C'$. Thus, any residual paths from $C$ to $C'$ in $P + H$ must cross $H$. After step 2, such residual paths cannot exist. Note that $K_1$ stays saturated after step 2. We now have a new saturated cut $K_2$ in $P$ with $C$ on one side and $C'$ and $C_H$ on the other side. After step 3, $K_2$ stays saturated (since $C'$ and $C_H$ are on the same side of $K_2$) and we get another saturated cut $K_3$ in $P$ separating $C_H$ and $C'$.

Consider an augmenting path implementation of the flow fixing procedure in step 4. Each augmenting path has both its endpoints on $C_H$ so it crosses neither $K_2$ nor $K_3$ (as it would have to cross $K_2$ or $K_3$ in both directions). Hence, $K_2$ and $K_3$ stay saturated after step 4 so there is no residual path in $P + H$ from $C$ to $C'$, from $C$ to $C_H$, or from $C_H$ to $C'$. The flow fixing procedure ensures that there are no residual paths in $P + H$ from excess vertices on $C_H$ to deficit vertices on $C_H$. This shows the desired. □

**Generalizing PUSHFLOWTHROUGHHOLES to $k$ additional holes** Now let us generalize Algorithm PUSHFLOWTHROUGHHOLES($P, C, C', H$) to arbitrary $k$. Instead of a single hole $H$, it gets a set $\{H_1, \ldots, H_k\}$ of holes as its fourth parameter. For $k \geq 2$, we can regard $P + H_1 + \ldots + H_{k-1}$ as a piece $P'$ with one hole $H_k$ in addition to the two bounded by $C$ and $C'$. Hence the call PUSHFLOWTHROUGHHOLES($P, C, C', \{H_1, \ldots, H_k\}$) solves the problem. Plugging in these parameters in the procedure for the single hole case, we get the following recursive algorithm:

Algorithm PUSHFLOWTHROUGHHOLES($P, C, C', \{H_1, \ldots, H_k\}$):

1. Execute PUSHFLOWTHROUGHHOLES($P, C, C', \{H_1, \ldots, H_{k-1}\}$) (finds max flow in $P'$ from $C$ to $C'$).

2. Execute PUSHFLOWTHROUGHHOLES($P, C, C_k, \{H_1, \ldots, H_{k-1}\}$). Denote the value of this flow by $d_2$ (finds max flow in $P'$ from $C$ to $C_k$).

3. Execute PUSHFLOWTHROUGHHOLES($P, C_k, C', \{H_1, \ldots, H_{k-1}\}$). Denote the value of this flow by $d_3$ (finds max flow in $P'$ from $C_k$ to $C'$).

4. Run a flow fixing procedure along $C_k$ in $P + H_1 + \ldots + H_k$ (i.e., in $P' + H_k$).

5. Let $\ell_2$ be the sum of excesses over vertices of $C_k$ with an excess, and let $-\ell_3$ be the sum of deficits over all vertices of $C_k$ with a deficit; reset the flow to zero in $P' + H_k$, and rerun the previous steps with a limits of $d_2 - \ell_2$ and $d_3 - \ell_3$ on the flow pushed in steps 2 and 3, respectively.

**Lemma 8.** *A call to* PUSHFLOWTHROUGHHOLES($P, C, C', \{H_1, \ldots, H_k\}$) *computes a max flow from $C$ to $C'$ in $P + H_1 + \ldots + H_k$. The total number max flow computations in $P$ between boundaries of holes and the total number of calls to a flow fixing procedure is $O(1)$ (for constant $k$).*

*Proof.* Correctness follows from Lemma 7 using induction on the recursive procedure. The number of max flow computations in $P$ and the number of calls to the flow fixing procedure are both exponential in $k$. Since $k = O(1)$, the second part of the lemma follows. □

# 4 An Efficient Implementation

In this section, we give an efficient implementation of Algorithm MaxPreflowToBoundaries. Recall that this algorithm applies Algorithm MaxPreflowInSeparatedGraph to obtain a max preflow from $s$ to an outer boundary $C'$, outside of $C'$, given a preflow from $s$ to $C$, outside of $C$, where $C$ is the parent of $C'$ in $\mathcal{F}$. We show that by maintaining flow assignments implicitly, such a call to Algorithm MaxPreflowInSeparatedGraph can be implemented to run in $O((|P| + |\partial P|^2) \log^2 |P|)$ time, where $P$ is the outer piece of $C'$. Combined with Lemma 2, this gives us the required $O(n \log^3 n)$ time bound for Algorithm MaxPreflowToBoundaries. An important tool that we use is the fast implementation of the flow fixing procedure by Borradaile et al. [4] which we go through briefly in the following.

## 4.1 The Flow Fixing Procedure of Borradaile et al.

Recall that the flow fixing procedure gets as input a cycle separator $C$ for a graph $G = G_1 + G_2$ and a pseudoflow in $G$ where all excess and deficit vertices (that are not at sources or at sinks) are on $C$. The procedure updates the flow network by sending as much flow as possible, among vertices of $C$, from excess vertices to deficit vertices. The efficient implementation of the flow fixing procedure from [4] takes advantage of the fact that the vertices of $C$ are on a single simple cycle in the plane. It processes the vertices of $C$ one by one in cyclic order. At each step, if the current vertex $v_i$ has excess, as much flow as possible is sent from $v_i$ to the yet unprocessed vertices. Similarly, if $v_i$ has a deficit, as much flow as possible is sent from the unprocessed vertices to $v_i$. By processing the vertices one by one, the procedure maintains the invariant that no flow can be sent from a processed vertex of $C$ with excess to any other vertex of $C$, or to a processed vertex of $C$ with deficit from any other vertex of $C$, without violating the capacity constraint of the network.

An important tool for the fixing procedure is an algorithm of Hassin [19] for (single-source single-sink) max flow in a plane digraph where the source and sink are on the same face. This algorithm works by adding an infinite capacity arc $e_\infty$ from the sink to the source and then computing shortest path distances in the dual graph from the face to the right of $e_\infty$. These distances define a *potential function* over the face which induces a max flow in the primal graph. Specifically, the flow $f(e)$ on an arc $e$ is defined by the difference between the potential of the face to the left of $e$ and the face to the right of $e$. This flow assignment defines a circulation, and after removing $e_\infty$ from this circulation we get a max $st$-flow.

In the flow fixing procedure we connect the unprocessed vertices of $C$ with edges of infinite capacity in both direction, thereby essentially identifying all these vertices with $v_{i+1}$, the vertex that follows $v_i$ on $C$. Then we can use Hassin's algorithm to compute a max flow from $v_i$ to $v_{i+1}$ if $v_i$ has an excess, or from $v_{i+1}$ to $v_i$ if $v_i$ has a deficit. This way using $O(|C|)$ calls to Hassin's algorithm we send flow among the vertices of $C$ such that at termination there is no residual path from any excess vertex $C$ to any deficit vertex of $C$.

The fixing procedure uses infinite capacity arcs between consecutive vertices of $C$. There is an arc $e_\infty$ between $v_i$ and $v_{i+1}$ as well as infinite capacity arcs between two consecutive unprocessed vertices of $C$. The vertices of $C$ are connected by a cycle anyway, and we do not want to change the graph $G$ during the fixing procedure. Therefore, instead we change the capacities of the relevant arcs of $C$ to infinity when needed. Instead of removing arcs with infinity capacity, we change back the capacity of the arc to its previous value. Let $e$ be an arc of $C$ such that we changed the capacity of $e$ to infinity and back to its previous value. The potential function that we found using Hassin's algorithm defines a flow of some value on $e$. If $e$ is the arc $e_\infty$ between $v_i$ and $v_{i+1}$ then we should remove the flow assigned to $e$ in order to get the correct max $v_i v_{i+1}$-flow (or max $v_{i+1} v_i$-flow). In addition, since we should send as much flow as possible from $v_i$ to $v_{i+1}$ if $v_i$ has an excess, or in the opposite direction of $v_{i+1}$ has a deficit, we assign to $rev(e)$ the maximum amount of flow that its residual capacity permits. If $e$ is one of the arcs that connect unprocessed vertices of $C$ we should remove the flow assigned to $e$, so we will not violate its capacity. This way, the result of each iteration of the fixing procedure is a potential function for the faces of $G$ and a flow assignment for the arcs of $C$ which we should add or reduce from the circulation induced by the potential function.

During the course of the fixing procedure, we maintain the current flow among the vertices of $C$ by accumulating the flow assignment for arcs of $C$ and a potential function $\phi$ over the face of $G$, which defines a circulation on this graph. Every iteration uses the residual network for the flow found in all previous steps.

As we described, Hassin's algorithm finds a max flow by computing a shortest path tree in the dual. To describe this part further, let us define $X^*$ to be the set of faces incident to vertices of $C$. This set contains the dual vertices from which we start the shortest path computations. Since the vertices of $G$ have constant degree, we have that $|X^*| = O(|C|)$. Moreover, let $G^*_{-C}$ be the graph obtained from $G^*$ by removing the dual arcs of $C$ and their reverses. Note that this splits $G^*_{-C}$ into two disconnected parts, the inside and outside of $C$. Each iteration of the fixing procedure applies Hassin's algorithm, which compute shortest path distances from a of $X^*$ in $G^*$ with respect to a weight function on the arcs induced by the arc lengths $l(u, v)$ of $G^*$ and the current potential function $\phi$. More precisely, the *reduced length $l_\phi$ with respect to $\phi$* is used, where $l_\phi(u, v) = l(u, v) + \phi(u) - \phi(v)$ [23]. A crucial observation to make in order to get an efficient implementation is that the only reduced distances that we are interested in in $G^*$ are those starting and ending in vertices of $X^*$ and these can be obtained from the $X^*$-to-$X^*$ distances in $G^*_{-C}$ and the restriction of $\phi$ to $X^*$, together with the lengths of the arcs dual to arcs of $C$. To see this, consider a path $Q = v_1 v_2 \dots v_k$ in $G^*_{-C}$ with $v_1, v_k \in X^*$. Then by a telescoping sums argument, the reduced length $l_\phi(Q)$ of $Q$ is $\sum_{i=1}^{k-1} l(v_i v_{i+1}) - \phi(v_{i+1}) + \phi(v_i) = l(Q) + \phi(v_1) - \phi(v_k)$, showing the desired.

In order to compute shortest path distances in $G^*$, FR-Dijkstra is applied, where $G^*_{-C}$ is represented by a matrix of $X^*$-to-$X^*$ distances and the dual arcs of $C$ are given explicitly. The algorithm only computes distances among vertices belonging to $X^*$. As we saw above, these values suffice both to update the flow on $C$ and to update the $X^*$-to-$X^*$ reduced lengths in $G^*_{-C}$. Hence, it suffices for the algorithm to maintain only the *restriction* of the face potential to vertices of $X^*$.

After processing all vertices of $C$, we get a restriction of the potential function $\phi$ which defines the reduced lengths among vertices of $X^*$. We want to extend this potential function to all faces of $G$, so that it will define a circulation in $G$. This circulation, together with the flow on arcs of $C$ give us the desired flow among the vertices of $C$. Borradaile et al. [4] showed that using one call to FR-Dijkstra on $X^*$, followed by one call to Dijkstra's algorithm over the entire graph $G$, we can get the desired potential function for all faces of $G$. Once we have this function, it is easy to compute the circulation it induces and add the flow on arcs of $C$ to get the desired flow which balance the excesses and deficits at vertices of $C$ as much as possible.

## 4.2 Implementation of Algorithm MaxPreflowInSeparatedGraph

We are now ready to describe our implementation of Algorithm MaxPreflowInSeparatedGraph. Let $P$, $C$, and $C'$ be defined as in the beginning of the section and assume that a max preflow $f_{sC}$ from $s$ to $C$ has been computed. Denote by $H_C$ and $H_{C'}$ the outside of $C$ and $C'$, respectively. The algorithm gets the max $sC$-preflow outside of $C$, $f_{sC}$, as an input and combines this preflow it with a max $CC'$-flow inside $C$ and outside of $C'$, to get a max $sC'$-preflow. We are not interested in the assignment of flow outside of $C$. In fact, we cannot even represent this flow if we want to implement the algorithm withing the desired time bound. We only need a representation that allows us to run the efficient implementation of the fixing procedure described above. From the description of the fixing procedure above, we conclude that it is enough to have an *implicit representation* of the input preflow $f_{sC}$ which consists of two parts:

1. The distances among the faces adjacent to $C$ in $ext(P)$, that is $DDG^*(H_C)$, where the length of a dual edge is the residual capacity of the primal edge with respect to $f_{sC}$.

2. The excess at each vertex of $C$ with respect to $f_{sC}$.

This does not cover the assignment of flow to arcs of $C$. However all vertices of $C$ are sinks of $f_{sC}$, therefore we may assume that the arcs of $C$ do not carry any flow in $f_{sC}$. This information is enough to run the efficient implementation of the fixing procedure described above.

The output of Algorithm MAXPREFLOWINSEPARATEDGRAPH is a max $sC'$-preflow. If $C'$ is a single vertex (that is, a sink $t$), then we are only interested in value of the preflow. Otherwise, we use this output as an input for computing a max $sC''$-preflow, where $C''$ is a child of $C'$ in $\mathcal{F}$. Therefore, it is enough to use an implicit representation of this kind also for the max $sC'$-preflow that we find as an output.

For an outer boundary $C$ that is a root of a tree in $\mathcal{F}$ we need to produce an implicit representation of $f_{sC}$ from an explicit max $sC$-flow. We first find such a flow using the algorithm of Borradaile and Klein [3], and then we compute the dense distance dual graph $DDG^*(H_C)$ with respect to the residual capacity using the MSSP algorithm of Klein [27]. This is done in $O(n \log^2 n)$ time for all trees of $\mathcal{F}$.

Our goal is to implement Algorithm MAXPREFLOWINSEPARATEDGRAPH to run in $O((|P| + |\partial P|^2) \log^2 |P|)$ time. Let us first consider the simple case where the only holes of $P$ are $H_C$ and the hole bounded by $C'$. This is the case where we do not require Algorithm PUSHFLOWTHROUGHHOLES from Section 3.3. In this case, Algorithm MAXPREFLOWINSEPARATEDGRAPH simply sends a max flow from $C$ to $C'$ in $P$ and then runs the flow fixing procedure on $C$ outside $C'$.

The first part, sending a max flow $f_{CC'}$ from $C$ to $C'$ in $P$, can be done in $O(|P| \log |P|)$ time by a single call to the algorithm of Borradaile and Klein [3] in $P$ (after adding a super-source and a super-sink, again we do not connect the shared vertices of $C$ and $C'$, if any, to the super-sink). From the max $CC'$-flow we computed in $P$ we obtain $IDDG^*(P)$, the internal dense distance dual graph with distances representing the residual capacity. Using Klein's MSSP algorithm [27] it takes $O((|P| + |\partial P|^2) \log |P|)$ time to construct $IDDG^*(P)$.

We now have dense distance dual graph on $DDG^*(H_C)$ that represents $f_{sC}$, and the dense distance graph $IDDG^*(P)$ that represents $f_{CC'}$, together with the amount of excess (from $f_{sC}$) and deficit (from $f_{CC'}$) at each vertex of $C$. These dense distance dual graphs, together with the dual arcs of $C$, allow us to apply the flow fixing procedure of Borradaile et al. [4] to $C$ in the part the graph outside $C'$. However, in the last step of the flow fixing procedure, where the procedure extends the potential function defined over the face adjacent to $C$ to all faces outside $C'$, using a call to FR-Dijkstra and a call to Dijkstra's algorithm, we limit this extension only to $P$. On one hand, we cannot afford extending the flow to the entire subgraph outside $C'$ since it might be much bigger than $P$, on the other hand we do require to extend the flow to $P$, since we are interested in the flow that enters $C'$ which is not represented explicitly in the current representation. So instead of using Dijkstra's algorithm to extend the flow to the entire graph as is [4], we use another call to FR-Dijkstra, where the arcs dual to $P$ and the arcs dual to $C$ are represented explicitly, but $H_C$ is represented using $DDG^*(H_C)$. This allows us to run the flow fixing procedure and to extend the flow to $P$ in $O((|P| + |\partial P|^2) \log^2 |P|)$ time.

If there are no deficit vertices on $C$, we have the desired $s$-to-$C'$ preflow, represented explicitly in $P$ and implicitly outside $C$. Otherwise, we rerun the above steps with a limit on the amount of flow pushed in $P$ as described in step 4 of Algorithm MaxPreflowInSeparatedGraph. In fact, in this case we can skip the computation of the explicit representation of the flow in $P$ following the first call to the flow fixing procedure.

Finally, we obtain an implicit representation of the combined flow $f_{sC'}$ using the MSSP algorithm. The total time for all these steps is $O((|P| + |\partial P|^2) \log^2 |P|)$, as required.

We assumed above that $P$ has only two holes, $H_C$ and the hole bounded by $C'$. We now show how to handle the general case with $k$ additional holes $H_1, \ldots, H_k$ bounded by cycles $C_1, \ldots, C_k$, respectively. We execute step 1 of Algorithm MaxPreflowInSeparatedGraph using an efficient implementation of Algorithm PushFlowThroughHoles($P, C, C', \{H_1, \ldots, H_k\}$). It follows from the description of this algorithm and from Lemma 8, that we should support a sequence of $O(1)$ operations of the following two types:

1. Find a max flow in $P$ from a cycle $C_i$ to a cycle $C_j$ (the single additional hole case of Algorithm PushFlowThroughHoles),

2. Run the flow fixing procedure along $C_i$ in $P + H_1 + \ldots + H_i$, for some $1 \leq i \leq k$.

During the course of Algorithm PushFlowThroughHoles, we represent the current pseudoflow in a way similar to the one above. We represent the pseudoflow explicitly in $P$, including the arcs of $C_1, \ldots, C_k$, and implicitly for the arcs inside the holes $H_1, \ldots, H_k$, using a dense distance dual graph for these holes with respect to the current residual capacity and the balance of flow at vertices of $C_1, \ldots, C_k$. Before the initial call to Algorithm PushFlowThroughHoles, the flow in $H_1, \ldots, H_k$ is all zero, so we can represent the initial residual network in each $H_i$ implicitly using DDG$^*(H_i)$ with the distances defined by the original capacity function. By Lemma 6, all of these dense distance dual graphs can be precomputed within the desired overall time bound of our algorithm.

The first kind of operation, sending max flow from one cycle to another inside $P$ can be implement by simply using the algorithm of Borradaile and Klein [3] with a super-source and a super sink in $O(|P| \log |P|)$ time.

The second type of operation, running the fixing along $C_i$ in $P + H_1 + \ldots + H_i$ is done in a way similar to before, but this time we should also take into account the part of the graph in the other holes $H_1, \ldots, H_{i-1}$. Again, to apply the fixing procedure to $C_i$ we require the dual arcs of $C_i$ together with DDG$^*(H_i)$ and IDDG$^*(P)$. This time, in addition we also need the dual arcs of $C_j$ for every $1 \leq j \leq k$ and also DDG$^*(H_i)$ for every $1 \leq j \leq i-1$. All of these dense distance dual graphs, as well as the length of the dual arcs of $C_j$'s, are with respect to the current residual capacity. When we apply FR-Dijkstra during the fixing procedure, we do it over a graph that is defined by the union of all of these dense distance dual graphs and dual arcs of cycle separators. The total number of vertices remains $O(|\partial P|)$, so the running time of the flow fixing procedure remains $O((|P| + |\partial P|^2) \log^2 |P|)$ as before.

After each time we apply the fixing procedure, we update the residual capacity of the arcs dual to each cycle separator $C_j$ and keep the restriction of the potential function for each of the dense distance dual graph DDG($H_j$), so that it will represent correctly the current residual capacity.

When we are done, we get the explicit flow of $P$ from the dense distance dual graphs and the flow assignment on the arcs of the cycles, and from them we get the implicit representation of the max $Ct$-flow in the same way as we described before, without computing the explicit flow for arcs that are not in $P$.

We have shown that the two types of operations required for finding the max $Ct$-flow take $O((|P| + |\partial P|^2) \log^2 |P|)$ time. By Lemma 8, there are only $O(1)$ of them during a call to

Algorithm PUSHFLOWTHROUGHHOLES, and hence step 1 of Algorithm MAXPREFLOWINSEP-ARATEDGRAPH runs in $O((|P|+|\partial P|^2)\log^2|P|)$ time. The fixing procedures at step 3 and step 4 are implemented in a similar way, within the same time bound.

We can now conclude this section with the main result of the paper.

**Theorem 1.** *Given a planar n-vertex digraph $G = (V, E)$ with a fixed source $s$ in $G$, the max st-flow values from $s$ to every sink $t \in V \setminus \{s\}$ can be computed in a total of $O(n \log^3 n)$ time.*

## 5 Reporting Min Cut Sets

The same ideas we used in Section 3 and Section 4 can also lead to an efficient algorithm for reporting minimum $st$-cut sets for a fixed $s$ and a given $t$. For simplicity we shall refer to cut-sets simply as *cuts*. The algorithm is based on a similar preprocessing scheme as the algorithm from sections 3 and 4. We use the duality between cut-sets in a planar graph and cycles in the dual planar graph. Let $K$ be an $st$-cut set, that is a set of arcs whose removal separates $s$ from $t$. Then, the dual arcs of the arcs of $K$ form a cycle which separates $s$ from $t$ and goes clockwise around $t$. In particular, if $f$ is a max $st$-flow, then a cycle $M$ which separates $s$ from $t$, and goes clockwise around $t$, whose length is equal to the value of $f$ (which we denote by $|f|$) is a dual of a min-cut. In this case, we call $M$ a *shortest st-separating cycle*. If we consider the dual of the residual graph (the lengths are given by residual capacities with respect to $f$), then the length of $M$ is 0, since it corresponds to a saturated $st$-cut. The following observation is the key for this step. Fix a source $s$, a cycle $C$, and two faces $x$, $y$ adjacent to $C$. Then, for every shortest $st$-separating cycle $M$ (for any sink $t$), which contains a subpath $Q$ from $x$ to $y$ that is embedded outside $C$, the path $Q$ is one of two possible paths: either the shortest $x$ to $y$ path that goes clockwise around $s$, or the shortest $x$ to $y$ path that goes counterclockwise around $s$, regardless of the choice of $t$. Our algorithm, after some preprocessing, is capable of reporting the arcs of the shortest $st$ separating cycle in time that is roughly proportional to its length.

We begin by giving a simple characterization of $st$-cuts. For this characterization we assume that for every sink $t$, the value of the min $st$-cut is positive. It is easy to identify sinks that satisfy this condition, since these are the vertices reachable from $s$ after removing all zero-capacity arcs.

**Lemma 9.** *Let $f$ be an st-flow in $G$. Let $M$ be a cycle of length 0 in $G_f^*$, the dual of the residual network for $f$. Then, if there is any flow crossing $M$, or in other words if there is an arc $e$ of $G$ such that the dual arc of $e$ is in $M$ and $f(e) \neq 0$, then $M$ is a shortest st-separating cycle. On the other hand, a shortest st-separating cycle $M$ has length 0 in $G_f^*$ and there is some flow crossing $M$. Furthermore, for every arc $e$ of $G$ whose dual in $M$, $f(e) \geq 0$.*

*Proof.* Let $M$ be a cycle of length 0 in $G_f^*$, $M$ separates the vertex set of $G$ into two sets – $S$ and $T$. Let $S$ and $T$ be such that every $e = (u, v)$ whose dual is in $M$ satisfies $u \in S$ and $v \in T$. Since the total length of $M$ is zero, we get that the residual capacity of every arc $e$ whose dual is in $M$ is zero. Therefore, for every such $e$, we get $f(e) \geq 0$. If there is flow crossing $M$, then the net amount of flow going from $S$ to $T$ is positive. Hence, it has to hold that $s \in S$ and $t \in T$, which means that $M$ corresponds to an $st$-cut. The residual capacity of $M$ is zero, so it is a shortest $st$-separating cycle.

The other direction follows from the discussion above, since a shortest $st$-separating cycle corresponds to a saturated $st$-cut. □

We use an algorithm similar to the one from the previous sections. However, this time finding a maximum preflow is not enough, since a max $st$-preflow might have a saturated cut whose value is greater than the value of the max $st$-flow. Therefore, this time we implicitly compute a max $st$-flow.

In contrary to the previous sections, where we guided our flow partition using a recursive decomposition, this time we use an $r$-division of $G$ as defined in [15]. It is a decomposition of $G$ into $O(\frac{n}{r})$ pieces of size $O(r)$. Each piece has $O(\sqrt{r})$ boundary vertices and a constant number of holes. An $r$-division can be computed in $O(n \log r + (n/\sqrt{r}) \log n)$ time [21].

For every vertex $t$ in the same piece as $s$, we compute a max $st$-flow and find a min $st$-cut using the algorithm of Borradaile and Klein [3], for a total of $O(rn \log n)$ time. We store all these cuts explicitly using $O(rn)$ space. It is left to consider only the case where $s$ and $t$ are in different pieces.

First, for every piece $P$ in the $r$-division with outer boundary $C$, we compute a max $sC$-flow using the algorithm of Borradaile and Klein [3] (using a super-sink). We will use these max flows for our flow partition algorithm. The total time for these computations over all pieces is $O(n^2 \log n/r)$, and we produce implicit representations of these flows in total of $O(n^2 \log n/r)$ time using the MSSP algorithm of Klein [27]. We also compute the dense distance dual graph of all holes in the $r$-division. Each of them is computed with the MSSP algorithm in $O(n \log n)$ time, which results in $O(n^2 \log n/r)$ time for all holes.

Then, we apply Algorithm IMPLICITMAXFLOW which is described next for every sink $t$ that is not in the same piece $s$. We denote by $P$ the piece containing $t$, and let $C$ be the outer boundary of $P$. This algorithm finds a max flow $f_{st}$ from $s$ to $t$, the flow is represented explicitly in $P$ and implicitly in the rest of the graph.

Algorithm IMPLICITMAXFLOW($t$):

1. Let $f_1$ be the max $sC$-flow strictly outside of $C$ computed in advance (represented implicitly).

2. Compute max $Ct$-flow $f_2$ inside of $C$. The flow $f_2$ is represented explicitly in $P$, if there are additional holes inside $C$, then $f_2$ is computed using Algorithm PUSHFLOWTHROUGHHOLES and $f_2$ is represented only implicitly in these holes.

3. Combine $f_1$ and $f_2$, and run the flow fixing procedure along $C$.

4. Compute the value of the max $st$-flow, denoted by $d$ as follows. Let $-\ell$ be the sum of deficits over all vertices of $C$ with deficit. Then $d := |f_2| - \ell$.

5. Restore the flow $f_1$ from step 1.

6. Push $|f_1| - d$ units of flow from vertices of $C$ with excess back to $s$ in the residual network of $f_1$ (strictly outside of $C$), thus obtaining an $sC$ flow $f_1'$ of value $d$ (represented implicitly).

7. Compute a max $Ct$-flow $f_2'$ inside $C$ with a limit of $d$ units on its value.

8. Combine the flows $f_1'$ and $f_2'$ and run the flow fixing procedure along $C$ into a single flow $f_{st}$, represented explicitly in $P$ and implicitly in the rest of the graph.

For every $t$, step 2 takes $O(r \log^2 n)$ time using the max flow algorithm of Borradaile and Klein and Algorithm PUSHFLOWTHROUGHHOLES. Step 7 is similar. In step 3 and step 8 we run the flow fixing procedure, which requires $O(r \log^2 n)$ as well. It remains only to bound the time for step 6.

In step 6, we return excess flow from $C$ to $s$, which is a standard step in the flow partition scheme that we use. Johnson and Venkatesan [24] implemented the flow return step explicitly, we cannot do this since it will take too much time. Moreover, in our case the flow $f_1$ is represented implicitly. The solution we used in Algorithm MAXPREFLOWINSEPARATEDGRAPH was to recompute the flow, in our case $f_1$, however we cannot afford spending the time required by this recomputation here. Note that we compute $f_1$ only once for all vertices of $P$. In contrast, in step 7 we send back flow from $C$ to $t$ as in Algorithm MaxPreflowInSeparatedGraph, we can do it in this step since recomputing a flow from $t$ to $C$ is possible within our time bound. Next we describe how to push back flow from $C$ to $s$ in the implicit representation of $f_1$, obtaining an implicit representation of $f_1'$.

It makes the implementation of this step easier if we assume that every face that is adjacent to a vertex of $C$ is also adjacent to an edge of $C$. This assumption can be satisfied by vertex-splitting along all boundaries of the $r$-division, as we described in Section 2.2 to obtain a graph of constant degree. After the vertex-splitting, the graph $G$ is no longer triangulated, and the size of the boundaries might increase. However, the blow-up in the size of the boundaries is by a constant factor, since the degree of each vertex was constant also before the splitting. For a face $x$ adjacent to $C$ we denote by $e(x)$ the edge of $C$ that is adjacent to $x$.

We divide the $|f_1| - d$ units of flow, which we want to push back from $C$, among the vertices of $C$ in an arbitrary way, such that we will not return from any vertex amount of flow greater than the excess it has in $f_1$.

We show how the implicit representation of the flow outside $C$ changes, when we push flow from vertices of $C$ back to $s$, in the residual network of $f_1$. Since the flow $f_1$ goes from $s$ to $C$, the flow is pushed back along paths that do not contain any vertex of $C$, except their starting vertices. Recall that the implicit representation contains all pair distances between faces adjacent to vertices of $C$ in the dual of the residual graph for $f_1$. We first show the change in this implicit representation after pushing back $k$ units of flow from some vertex $v$ of $C$ to $s$. As we show below, we get the same implicit representation regardless of the exact path $R_v$ along which the flow is pushed back.

As in Section 4, we denote by $H_C$ the hole of $P$ outside $C$ and $\text{DDG}^*(H_C)$ is the dense distance graph representing the residual capacity in this hole. Consider an arc $(x, y)$ in $\text{DDG}^*(H_C)$ that connects two faces adjacent to vertices of $C$. It corresponds to a dual path $Q$ in the dual of $\text{ext}(P)$, embedded strictly outside $C$. Assume that we push $k$ units of flow along a path $R_v$ from a vertex $v \in C$ to $s$.

We say that a directed dual path $M$ in the dual graph crosses a directed path $N$ in the primal graph if $M$ contains a dual of an arc of $N$ or its reverse. In the former case, we say that *M crosses N from the right side*, whereas in the latter we say that *M crosses N from the left side*. If $Q$ crosses $R_v$ exactly once, then after pushing $k$ units of flow along $R_v$ –

1. if $Q$ crosses $R_v$ from the right side, that is it contains a dual of an arc of $R_v$, then the length of $Q$ decreases by $k$,

2. if $Q$ crosses $R_v$ from the left side, that is it contains a reverse of a dual of an arc of $R_v$, then the length of $Q$ increases by $k$.

This can be further generalized. Consider the order of the crossings between $Q$ and $R_v$ as we encounter them when we traverse $R_v$. Consecutive crossings alternately increase and decrease the length of $Q$ by $k$. If the total number of crossing is odd, then the length of $Q$ changes by $+k$ or $-k$, depending on the direction of the first crossing. On the other hand, if the number of crossings is even, then the length of $Q$ remains unchanged (see Figure 4).

Observe that $Q$ splits $\text{ext}(P)$ into two regions $E_1$ and $E_2$. Let $E_1$ be the region containing $s$. Formally speaking, $Q$ does not start at $C$, but at two faces that are adjacent to vertices of $C$. We divide the graph outside of $C$ with a curve that is composed of the embedding of $Q$, as well as two curves that connects $x$ and $y$ to $e(x)$ and $e(y)$ respectively. In particular, the cycle $C$ itself is divided into two parts $C_1$ and $C_2$, where $C_i$ belongs to $E_i$.

In the beginning of our algorithm, we computed $\text{DDG}^*(H_C)$ that represents the flow $f_1$ outside $C$. We now extend this computation to make it possible to update the length of $Q$. Fix

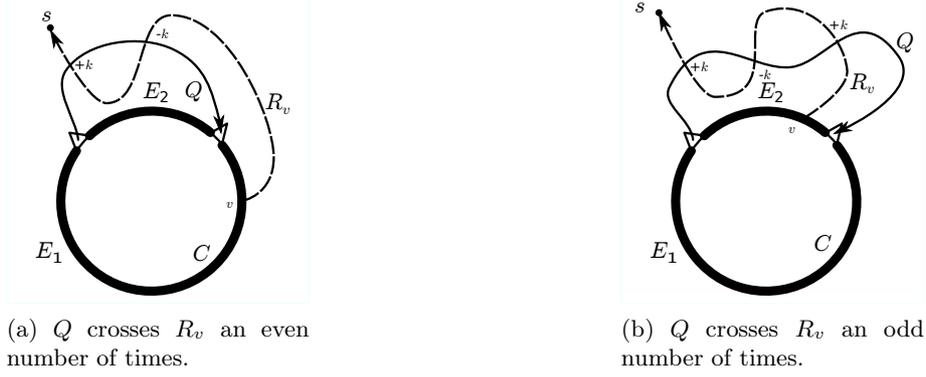

(a) $Q$ crosses $R_v$ an even number of times.

(b) $Q$ crosses $R_v$ an odd number of times.

Figure 4: Illustration of lemmas 11 and 12. The cycle $C$ is marked in bold, $Q$ is solid and $R_v$ is marked with a dashed line. If $k$ units of flow are pushed along $R_v$, then every crossing between $Q$ and $R_v$ contributes a change of $+k$ or $-k$, alternately, in the length of $Q$.

an arbitrary vertex $w$ on $C$ an let $R'$ be an arbitrary the path connecting $w$ to $s$ outside of $C$. Instead of computing $\mathrm{DDG}^*(H_C)$, we compute for each pair of faces $x$, $y$ that are adjacent to vertices of $C$ and lie outside $C$ two shortest paths connecting them. One of them is the shortest dual path that crosses $R'$ an even number of times, while the other is that shortest dual path that crosses $R'$ an odd number of times. It is clear that the shortest path from $x$ to $y$, whose length defines the length of $(x,y)$ in $\mathrm{DDG}^*(H_C)$, is always one of these two dual paths. We show that the lengths of such $x$-to-$y$ dual paths can be efficiently updated after some flow is pushed via paths from $C$ to $s$.

**Lemma 10.** *Let $Q$ be a dual path from $x$ to $y$ and let $R_v$ be a path connecting $v \in C$ and $s$, both strictly outside $C$.*

*If $v$ is in the same region as $w$, the vertex which defines $R'$, then the parity of the number of crossings of $Q$ with $R'$ is the same as the parity of the number of crossings of $Q$ with $R_v$. Otherwise, if $v$ and $w$ are in different regions, then these parities are different.*

*Proof.* We claim that the parity of the number of crosses of $Q$ with $R_v$ depends solely on whether $v$ is in $C_1$ or $C_2$. Indeed, in every time $R_v$ crosses $Q$, it switches from $E_1$ to $E_2$ or vice versa. From this, we get the desired. □

From this lemma we infer that for every $v \in C_1$ the number of crossing between the path $R_v$ and the dual path $Q$ is even, and for $v \in C_2$ the number of crossing is odd.

**Lemma 11.** *Let $Q$ be a dual path from $x$ to $y$ strictly outside $C$. Assume that we push $k$ units of flow from $v \in C$ to $s$ along a path $R_v$. Assume also that $Q$ crosses $R_v$ an even number of times (see Figure 4(a)). Then, pushing the flow along $R_v$ does not affect the length of $Q$.*

*Proof.* Every time $R_v$ crosses $Q$, it switches from $E_1$ to $E_2$ or vice versa. Hence, $Q$ alternates between dual arcs of $R_v$ and reverse of dual arcs of $R_v$. A crossing in one direction decreases the length of $Q$ by $k$ while a crossing in the other direction increases the length of $Q$ by $k$. The number of crossing of each type is the same, so the length of $Q$ remains intact. □

The following lemma considers the clockwise order of $x$, $v$ and $y$ around $C$. The faces $x$ and $y$ do not belong to $C$, but their order around this cycle, relative to $v$, is well-defined by $e(x)$ and $e(y)$.

**Lemma 12.** *Consider the scenario from the previous lemma, but assume that $Q$ crosses $R_v$ an odd number of times (see Figure 4(b)). If $x$, $v$ and $y$ are in clockwise order around $C$ then after pushing the flow along $R_v$, the length of $Q$ increases by $k$. Otherwise, if the order of $x$, $v$ and $y$ is counterclockwise, the length of $Q$ decreases by $k$.*

*Proof.* We consider the case, when $x$, $v$ and $y$ lie in clockwise order around $C$. The other case is symmetric.

In this case the crossing of $Q$ and $R_v$ which is closest to $v$ in $R_v$ is from the left side of $R_v$. The following crossings alternate between crossings from the right side and crossing from the left side. This implies that $Q$ crosses $R_v$ from the left side $t+1$ times and from the right side $t$ times, for some $t \geq 0$. Therefore, the length of $Q$ increases by $k$. □

From the above lemmas, we infer that when $k$ units of flow are pushed from $v \in C$ to $s$, the length of any path connecting faces adjacent to vertices of $C$ in the dual of $\text{ext}(P)$ changes by $-k$, $0$ or $k$. The actual path $R_v$ on which we push the flow is not important, the amount of change in the length of a path $Q$ is determined only by $k$, by the parity of the number of crossings between $Q$ and $R'$, whose choice was arbitrary and independent of $v$, and by the cyclic order of $x$, $v$, and $y$.

Moreover, the total amount of change in step 6 in the length of $Q$ is determined only by two factors. The absolute value of the change is the total amount of excess that we are returning from $C_1$, it is easy to obtain this value for every pair of dual vertices $x$, $y$ in total of $O(|C|^2) = O(r)$ time using dynamic programing. The sign of the value is determined by the cyclic order of $x$, $y$ and $C_1$. Note also that all $x$-to-$y$ paths with a give parity of number of crossing with $R'$ change by the same amount, so the shortest paths remain the same.

We update the lengths of all shortest paths connecting faces adjacent to vertices of $C$ and crossing $R'$ an even number of time or odd number of times. Then, for each pair of these faces we find smaller of the two distances we have, this way we obtain $\text{DDG}^*(H_C)$ for representing the $sC$-flow $f'_1$ of value $d$. Thus, step 6 can be implemented in $O(|C|^2) = O(r)$ time, since this is the number of entries in $\text{DDG}^*(H_C)$.

Determining which side of $x$ and $y$ is $C_1$ is easy, according to the parity of the number of crossing between $Q$ and $R'$. It remains to show how to compute the shortest paths among faces adjacent to vertices of $C$ whose number of crossing with $R'$ has a given parity. For this purpose, we use a procedure similar to [8, 26]. We build a new graph $G^{R'}$. First, we make an *incision* in $\text{ext}(P)$ along $R'$, thus obtaining a graph $\bar{G}^{R'}$, in which there are two copies of $R'$: $R_1$ and $R_2$. Note that the starting and ending vertices of $R'$ are also split. The graph $G^{R'}$ is obtained by taking two copies of $\bar{G}^{R'}$ and identifying the vertices of $R_1$ in each of the copies of the graph with the vertices of $R_2$ in the other copy of the graph. Each vertex $v$ from the original graph has two copies $v'$ and $v''$ in $G^{R'}$. Observe that a path from $u'$ to $v''$ corresponds to a path that connects $u$ with $v$ in $\text{ext}(P)$ and crosses $R'$ an odd number of times. Similarly, a path from $u'$ to $v'$ (or from $u''$ to $v''$) can be mapped to a path crossing $R'$ an even number of times. The same is true also if we consider the dual graph of $G^{R'}$, every dual vertex has two copies such that a path from one copy to the other corresponds to a dual path in the original graph that crosses $R'$ and odd number of times, and a path from a copy back to itself corresponds to a dual path in the original graph that crosses $R'$ an even number of times. This means that by using the MSSP algorithm of Klein [27] from one of the copies of the dual vertices corresponding to the boundary faces along $C$ we can compute the length of the shortest dual paths connecting these dual vertices and crossing $R'$ an even or an odd number of times. We uses the length defined by the residual capacities with respect to $f_1$. This takes $O((n + |C^2|) \log n) = O(n \log n)$ time for all pairs of dual vertices around $C$. Note that it is enough to compute these paths only once for each piece $P$, before we apply Algorithm IMPLICITMAXFLOW.

Summing over all steps of the algorithm, we get $O(n \log n)$ preprocessing time per piece, and in addition $O(r \log^2 n)$ running time per sink. Therefore, we obtain the following:

**Lemma 13.** *Using Algorithm* IMPLICITMAXFLOW *we can find an implicit representation of a max st-flow, $f_{st}$, for every vertex $t$ in a piece different from $s$, such that $f_{st}$ is represented explicitly in the piece containing $t$ and implicitly for the rest of the graph, in $O(n^2 \log n/r + nr \log^2 n)$ total time.*

We now show how to use the flow $f_{st}$ that we found above to store implicitly a min $st$-cut. First note that there are two special cases where this is not required. We already mentioned that we explicitly store a min $st$-cut for every $t$ in the same piece as $s$ using a total of $O(nr)$ space. Another case where we store the $st$-cut explicitly is when we find that the value $f_{st}$, for some sink $t$, is the same as the value of $f_1$, the maximum flow from $s$ to the outer boundary of the piece containing $t$. In this case, a min $sC$-cut is also a min $st$-cut, there are only $O(n/r)$ different such cuts, and we store all of them explicitly in $O(n^2/r)$ space.

In the remaining case, there is a min $st$-cut that contains at least one arc of $P$. We show below how to use Lemma 9 to find the cut, which is dual to a separating cycle in the dual planar graph. We store the arcs of the cut inside $P$ explicitly, and the arcs of the cut that are not in $P$ implicitly, by storing arcs of the dense distance graphs of holes of $P$ that corresponds to these parts of the cut. This requires $O(r)$ space per sink and a total of $O(nr)$ space. We also keep all the MSSP data structures that were used to compute the dense distance dual graph, this takes a total of $O(n^2/r)$ space. From an MSSP data structure, we can extract a shortest path of length $\ell$ between two vertices in $O(\ell \log \log n)$ time (recall that each arc of a dense distance dual graph is defined by a shortest path that is stored in one of the MSSP structures). Note that even though the distances in the dual graph that defines the MSSP data structure are changing during the execution of the algorithm, as the residual capacity changes, the paths themselves do not change. One special case is arc of the dense distance graph that represent the flow in $\text{ext}(P)$ – in this case the arc corresponds to a shortest path in one of two possible MSSP structures (see the details of step 6 of Algorithm IMPLICITMAXFLOW), we need to store both structures and for each arc remember from which of the two it was originated.

It remains to show how to find the arcs that define the cut. This step is also done during the preprocessing of the data structure. Consider the representation of $f_{st}$, that is an explicit flow assignment for arcs of $P$ and dense distance dual graphs for the holes of $P$, including $\text{ext}(P)$, were the distances are defined by the residual capacity with respect to $f_{st}$. First, remove all arcs of positive length, both in $P$ and in the dense distance dual graphs, since they cannot be in a saturated cut. By Lemma 9 we need to find a cycle such that at least one of the arcs of the cycle carries flow. For arcs of $P$, we have the explicit representation of the flow, so it trivial to determine if an arc carries a flow. Similarly, for each arc $(x, y)$ in a dense distance dual graph of a hole of $P$, we need to check whether it corresponds to a path that is crossed by some flow. Since every arc of the minimum cut has non-negative flow assigned to it, it is enough to compare the length of $(x, y)$ defined by the residual capacities with respect to $f_{st}$ with the length of this arc for defined by the original capacity function.

Thus, the problem can now be stated the following way: given a directed graph with a set of distinguished arcs (those who carry flow), find a cycle with at least one distinguished arc. We solve this problem in linear time, by first finding strongly connected components and then finding a strongly connected component with a distinguished arc $(x, y)$. It follows that there has to be a $yx$ path in this strongly connected component.

The cycle we find consists both of dual arcs of $P$ and arcs from dense distance dual graphs. We find the cycle in $O(r)$ time per sink, and store it using $O(r)$ space. As we mentioned above, on query time we recover the paths inside the holes of $P$ that correspond to these arcs from the MSSP structures in total of $O(|M|\log\log n)$ time, where $M$ is the shortest $st$-separating cycle.

Recall that in the preliminaries we added some zero-capacity arcs to the graph to make it satisfy some structural properties (bounded face size, decomposition with connected pieces). Adding these arcs does not change the value of a min $st$-cut, but might change the size of the cut-set. We remove these artificial arcs, and replace all the MSSP data structures with equivalent data structures (over the same set of vertices, inside the same subgraph) that store shortest paths in the original dual graph. First note that this does not change the distances in the MSSP structures (since instead of a dual arc of length 0 we merge the two dual vertices it is adjacent to into one). Second, note that a set of dual vertices that are on a single dual face, remains on a single dual face also when we remove the artificial arcs (we might need to add some infinite length artificial arcs to the dual graph, if the boundary of that dual face becomes disconnected). Therefore, by storing the MSSP data structure over the original dual graph we can return the cut-sets in time almost proportional to their sizes in the original graph (up to the $\log\log n$ factor we get from using the MSSP data structure).

This concludes the description of the algorithm, and proves the correctness of the following theorem.

**Theorem 2.** *There exists a data structure that, given a planar $n$-vertex digraph $G = (V, E)$ with a fixed source $s$ in $G$ and a parameter $r \in [1, n]$, can report the min st-cut sets for any sink $t$. The data structure requires $O(n^2 \log n/r + nr\log^2 n)$ time for preprocessing and the min st-cut set $M$ is reported in $O(|M|\log\log n)$ time. The data structure requires $O(n^2/r + nr)$ space.*

To minimize the required storage space we set $r = n^{1/2}$ and get $O(n^{3/2}\log^2 n)$ preprocessing time and $O(n^{3/2})$ space. To minimize the preprocessing time we set $r = (n/\log n)^{1/2}$ and get $O(n^{3/2}\log^{3/2} n)$ preprocessing time and $O(n^{3/2}\log^{1/2} n)$ space.

If we want to improve the query time to $O(|M|)$, we can avoid using the MSSP data structure for the queries, and store the shortest paths trees it represents explicitly. The MSSP structures are defined over total of $O(n/\sqrt{r})$ boundary vertices, each is a member of a constant number of structures. We find the explicit shortest path trees from each boundary vertex in linear time per tree using the shortest path algorithm of Henzinger et al. [20]. Therefore, the required additional time is $O(n^2/\sqrt{r})$ and the additional storage space is $O(n^2/\sqrt{r})$ as well.

**Theorem 3.** *There exists a data structure that, given a planar $n$-vertex digraph $G = (V, E)$ with a fixed source $s$ in $G$ and a parameter $r \in [1, n]$, can report the min st-cut sets for any sink $t$. The data structure requires $O(n^2/\sqrt{r} + nr\log^2 n)$ time for preprocessing and the min st-cut set $M$ is reported in $O(|M|)$ time. The data structure requires $O(n^2/\sqrt{r} + nr)$ space.*

To minimize the required storage space we set $r = n^{2/3}$ and get $O(n^{5/3}\log^2 n)$ preprocessing time with $O(n^{5/3})$ space. To minimize the preprocessing time we set $r = (n/\log^2 n)^{2/3}$ and get $O(n^{5/3}\log^{2/3} n)$ preprocessing time with $O(n^{5/3}\log^{2/3} n)$ space.

## 6 Concluding Remarks

We gave an $O(n\log^3 n)$ time algorithm for the problem of finding max $st$-flow values for a fixed source $s$ and all sinks $t \in V \setminus \{s\}$ in an $n$-vertex planar digraph $G = (V, E)$. The previous best known solution was to perform $n-1$ executions of a single-source single-sink max flow algorithm, which gave an $O(n^2\log n)$ time bound with the algorithm of Borradaile and Klein [3].

An immediate corollary of our result is a near-quadratic time algorithm for finding max $st$-flow values for all source/sink pairs $(s, t)$. We showed that the number of distinct max $st$-flow values is quadratic in the worst case. Hence, our algorithm is optimal up to logarithmic factors.

Prior to this result, computing all max-flow values in a directed planar graph required up to $\Theta(n^2)$ max-flow computations. We conjecture that a similar improvement is possible for general directed graphs. Another interesting direction to pursue is to prove or disprove the existence of a data structure similar to the oracle in [6], which can answer queries for max $st$-flow values in constant time after $o(n^2)$ preprocessing. A related question is whether it is possible to improve the time needed to find max $st$-flow values for $k$ given input pairs $(s, t)$? Also, is it possible to remove the log-factors, i.e., compute all max-flow values in optimal $O(n^2)$ time? Finally, our algorithm computes all cut-sets $\mathcal{C}$ in a planar digraph in $O(\min\{n^{5/2}\log^{3/2} n + \sum_{C\in\mathcal{C}} |C|\log\log n, n^{8/3}\log^{2/3} n + \sum_{C\in\mathcal{C}} |C|\})$ time. Is it possible to improve this running time?